\documentclass[reprint,
amsmath,amssymb,11pt,reprint,aps,prl,showpacs,floatfix]{revtex4-2}
\usepackage{lineno}
\usepackage[english]{babel}
\usepackage{multirow}
\usepackage[thinspace]{SIunits}
\usepackage{xspace}
\usepackage{ulem}
\usepackage{soul}
\usepackage{graphicx}% Include figure files
\graphicspath{{figures//}} %allows to omit the path in the \incudegraphics command
\usepackage{dcolumn}% Align table columns on decimal point
\usepackage{bm}% bold math
\usepackage{color}
\usepackage{wrapfig}
\usepackage{hyperref}
\usepackage{array}
\usepackage{booktabs}
\usepackage{multirow}
\usepackage{amssymb}
\usepackage{adjustbox}
\usepackage{marvosym}
\hypersetup{
    unicode=false,          % non-Latin characters in Acrobat’s bookmarks
    pdftoolbar=true,        % show Acrobat’s toolbar?
    pdfmenubar=true,        % show Acrobat’s menu?
    pdffitwindow=false,     % window fit to page when opened
    pdfstartview={FitH},    % fits the width of the page to the window
    pdftitle={My title},    % title
    pdfauthor={Author},     % author
    pdfsubject={Subject},   % subject of the document
    pdfcreator={Creator},   % creator of the document
    pdfproducer={Producer}, % producer of the document
    pdfkeywords={keyword1} {key2} {key3}, % list of keywords
    pdfnewwindow=true,      % links in new window
    colorlinks=true,       % false: boxed links; true: colored links
    linkcolor=blue,          % color of internal links (change box color with linkbordercolor)
    citecolor=blue,        % color of links to bibliography
    filecolor=blue,      % color of file links
    urlcolor=blue,       % color of external links
    plainpages=false        % influences the page numbering
}

\newcommand{\etal}{\textit{et al.}\xspace}

\newcommand{\LNOf}{\mbox{La$_4$Ni$_3$O$_{10}$}\,\xspace}
\newcommand{\LNOt}{\mbox{La$_3$Ni$_2$O$_{7}$}\,\xspace}
\newcommand{\Alg}{$A_{\rm{1g}}$\,\xspace}

\newcommand{\Blg}{${B_{\rm{1g}}}$\xspace}
\newcommand{\BZg}{${B_{\rm{2g}}}$\xspace}

\newcommand{\wn}{\rm{cm}$^{-1}$\,}

\begin{document}

\title{\Large Contrasting Momentum-Selective Spin-Density-Wave Gaps in Bilayer and Trilayer Nickelates}

\author{Jun Shu$^{1,2,10}$,
Jun Shen$^{1}$\textrm{\Letter},
Xiaoxiang Zhou$^{3,10}$,
Yinghao Zhu$^{7}$,
Qingsong Wang$^{1}$,
Dengjing Wang$^{2}$,
Weihong He$^{3}$
Jie Yuan$^{4, 5}$,
Kui Jin$^{4, 5, 6}$,
Dawei Shen$^{3}$
Congcong Le$^{3}$,
Jun Zhao$^{7}$\textrm{\Letter},
Zengyi Du$^{3}$\textrm{\Letter},
Ge He$^{1}$\textrm{\Letter}, 
Donglai Feng$^{3}$\textrm{\Letter}}

\affiliation{
$^1$ School of Mechanical Engineering\mbox{,} Beijing Institute of Technology\mbox{,} Beijing 100081\mbox{,} China \\
$^2$ Department of Applied Physics\mbox{,} Wuhan University of Science and Technology\mbox{,} Wuhan 430081\mbox{,} China\\
$^3$ Hefei National Laboratory\mbox{,} and New Cornerstone Science Laboratory\mbox{,} Hefei, Anhui 230088\mbox{,} China\\ 
$^4$ Beijing National Laboratory for Condensed Matter Physics\mbox{,} Institute of Physics\mbox{,} Chinese Academy of Sciences\mbox{,} Beijing 100190\mbox{,} China\\
$^5$ School of Physical Sciences\mbox{,} University of Chinese Academy of Sciences\mbox{,} Beijing 100049\mbox{,} China\\
$^6$ Songshan Lake Materials Laboratory\mbox{,} Dongguan\mbox{,} Guangdong 523808\mbox{,} China\\
$^7$ State Key Laboratory of Surface Physics and Department of Physics, Fudan University, Shanghai 200433, China\\
$^{10}$ These authors contributed equally: Jun Shu and Xiaoxiang Zhou\\
\textrm{\Letter} e-mail: 
jshen@bit.edu.cn;
zhaoj@fudan.edu.cn
duzengyi@hfnl.cn;
ge.he@bit.edu.cn;
dlfeng@hfnl.cn
}
\begin{abstract}
\section{Abstract}

Resolving where the density-wave gap opens in momentum space is essential for identifying the microscopic origin of the instability in layered nickelates. Using polarization-resolved electronic Raman scattering, we map the momentum selectivity of the spin-density-wave (SDW) gap in trilayer \LNOf. We observe a SDW-induced redistribution of spectral weight on both the $\alpha$ pocket at the Brillouin-zone centre and a portion of the $\beta$ pocket near the zone boundary, characterized by gap energies of approximately 55~meV.
In contrast, no comparable spectral weight suppression is observed along the diagonal region of $\beta$ pockets, implying little or no gap opening. This gap topology contrasts sharply with that in \LNOt, where anisotropic SDW gaps open solely on the $\beta$ pocket. Our results establish a distinct momentum-space gap topology between bilayer and trilayer nickelates, placing new constraints on the ordering wave vector and the mechanism of the density-wave instability relevant to superconductivity.

\end{abstract}
\maketitle

\section{Introduction}

Layered nickelate superconductors, including \LNOf and \LNOt, exhibit markedly different superconducting transition temperatures despite closely related crystal structures and electronic band topologies~\cite{Sun:2023,Zhu:2024,Wang:2024,Zhang2025PRX.15.021005,Zhang2025PRX.15.021008,Shi2025,Li:2025nature}. Both compounds host a density-wave (DW) instability in the normal state at $T_{\mathrm{DW}}\approx 140–150$~K~\cite{RIXS:2024,WutaoNMR:2025,Yanglexian:2025,RuiZhou:2025,Khasanov:2025,Kakoi2024JPSJ.93.053702,Wang:2024,ARPESYanglexian：2024,Damascelli:2025,Liu2022,RuiZhou:2025,Liuzhe:2024,Yuxiaohui:2024,GeHe:2025,Zhang2020NC,WangNanlin:2025,Lihaoxiang2017NC.8.704,Du2025arXiv.2405.19853v1,Gim:2025arXiv,Suthar:2025arXiv,Limingzhe:2025,Yang2024PRB.109.L220506,Zhang2024PRB.110.L180501,Zhang2025PRB.111.144502}, widely discussed in terms of spin- and/or charge-density-wave (SDW/CDW) formation. This ordering substantially reshapes the low-energy electronic structure, making the momentum dependence of the associated gap a key ingredient for understanding how superconductivity emerges.

Rather than providing a unified picture, existing experiments probe complementary but incomplete aspects of the DW state in \LNOf. X-ray scattering and optical measurements~\cite{Zhang2020NC,WangNanlin:2025} unambiguously establish the presence of DW order and a finite gap scale, respectively, but lack the combined energy and momentum resolution required to determine where the gap opens in reciprocal space. Momentum-resolved probes, such as angle-resolved photoemission spectroscopy (ARPES) and electronic Raman scattering~\cite{Lihaoxiang2017NC.8.704,Du2025arXiv.2405.19853v1,Gim:2025arXiv,Suthar:2025arXiv,Deswai:2025APL}, access this information more directly but have so far yielded mutually inconsistent conclusions regarding the involved Fermi-surface pockets. Real-space imaging by scanning tunneling microscopy (STM) further reveals a DW modulation and constrains the ordering wave vector, but does not directly resolve the momentum-dependent gap structure~\cite{Limingzhe:2025}. As a result, the topology of the DW gap in momentum space remains unresolved.

Symmetry-resolved electronic Raman scattering is a powerful probe of momentum-dependent gap formation and has been successfully applied to reveal both DW gaps and superconducting gap openings in cuprates and iron-based superconductors \cite{Devereaux:2007, Lazarevic:2020}. Our previous polarization-resolved electronic Raman scattering study on \LNOt\ demonstrated that the SDW gap opens in a highly momentum-selective manner~\cite{GeHe:2025}. In that system, an anisotropic SDW gap with strong-coupling character develops on the $\beta$ pocket, while the $\alpha$ pocket remains ungapped. This momentum-selective SDW gap structure provided a consistent framework for reconciling the seemingly disparate results reported by ARPES~\cite{Zhouxingjiang:2024,ARPESYanglexian：2024,ZXSHen:2025,Damascelli:2025}, optical conductivity~\cite{Liuzhe:2024,Yuxiaohui:2024}. 

\begin{figure}[ht]    \includegraphics[width=0.5\textwidth]{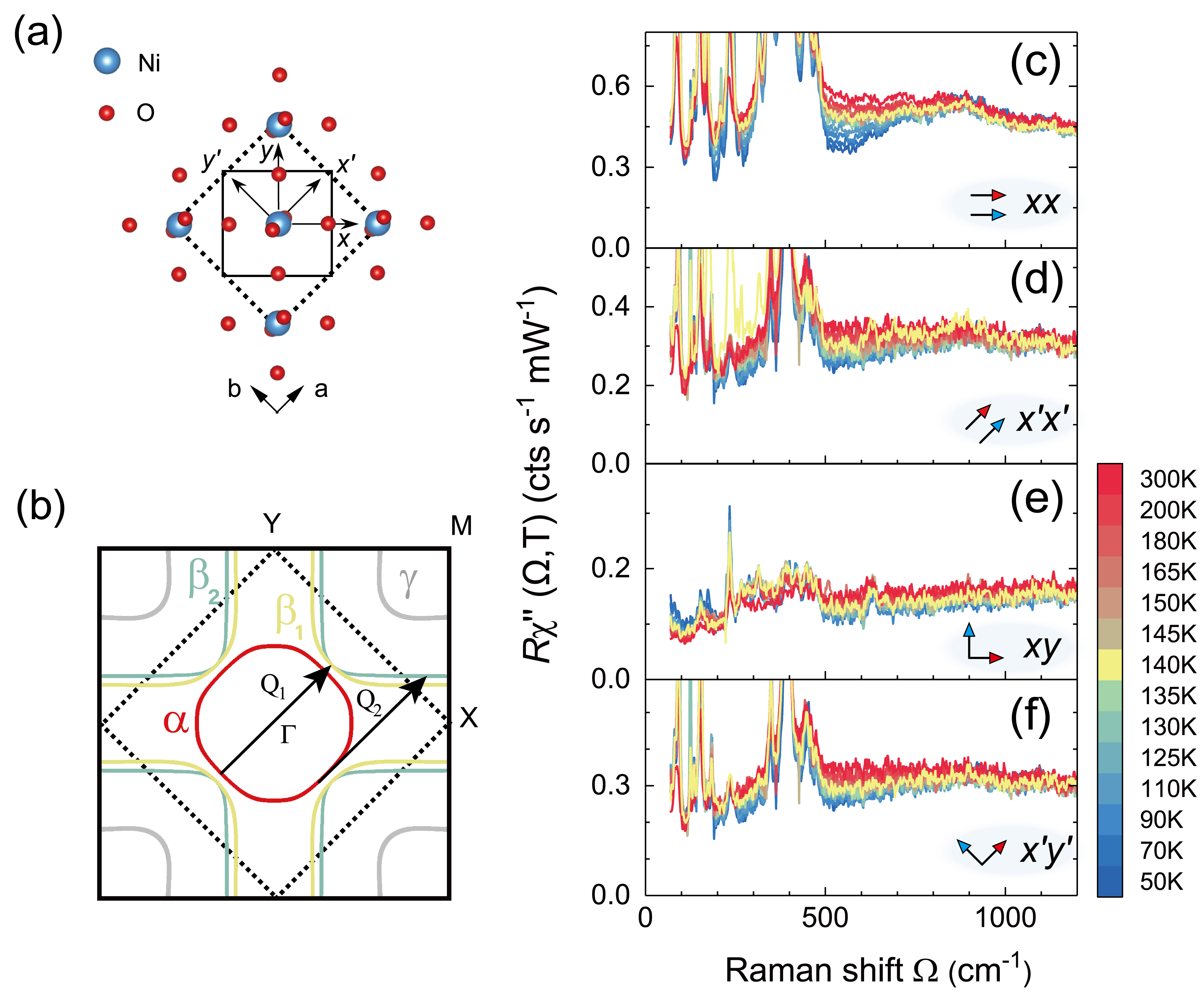}
   \caption{\textbf{Definition of the Ni–O unit cell, photon polarizations, Fermi pockets, and Raman responses in different polarization configurations.} 
(a) Solid (dashed) lines indicate the 1-Ni (2-Ni) unit cell. The $x$ and $y$ axes are aligned along the Ni–O–Ni bond directions, while the $x'$ and $y'$ polarizations are rotated by 45$^\circ$ clockwise, corresponding to the Ni–Ni directions. 
(b) Schematic Fermi surface obtained from a tight-binding model with $\alpha$ (red), $\beta_1$ (yellow), $\beta_2$ (green), and $\gamma$ (grey) pockets (see Supplementary Materials F for details). Black arrows denote candidate scattering wave vectors: \textbf{Q}$_1$ and \textbf{Q}$_2$. 
(c–f) Temperature dependence of the electronic continuum measured in the $xx$, $x'x'$, $xy$, and $x'y'$ channels.
}

   \label{fig:motivation}
\end{figure}

In this work, we use polarization- and symmetry-resolved electronic Raman scattering to directly determine the momentum-selective structure of the SDW gap in \LNOf. By comparing with the SDW gap topology previously established in \LNOt~\cite{GeHe:2025}, we uncover qualitative differences in both the momentum dependence and coherence of the SDW state. In \LNOf, Raman spectra reveal a pronounced SDW-induced spectral weight redistribution near the Brillouin-zone (BZ) center ($\alpha$ pocket) and boundary ($\beta$ pockets), corresponding to a characteristic gap scale of $\sim$55~meV, while no comparable suppression is observed along the $\Gamma$-M direction of the $\beta$ pockets, contrasting sharply with \LNOt, where strong-coupling SDW gaps open on the $\beta$ pocket. These observations indicate that the incommensurate SDW wave vector in \LNOf\ predominantly connects the $\alpha$ and $\beta$ pockets near the X/Y points ($Q_2$ in Fig.~\ref{fig:motivation}\textbf{b}), rather than the $\alpha$–$\beta$ nesting along the $\Gamma$-M direction ($Q_1$). Our results establish a distinct momentum-space SDW gap topology in \LNOf, providing new constraints on the microscopic origin of density-wave order in layered nickelates.

%In this work, we apply this approach to \LNOf and directly determine the momentum-selective structure of its SDW gap. By contrasting the SDW gap topology in \LNOf with that in \LNOt, we uncover qualitative differences in the momentum dependence and coherence of the SDW state, providing new constraints on the microscopic origin of density-wave order in layered nickelates. 

%In this study, we performed polarization-resolved electronic Raman scattering on \LNOf and resolved SDW-related spectral weight redistribution at the center and boundary of the BZ, with a characteristic gap size of $\approx$ 55 meV, while no spectral weight changes were observed near the $(\pm\pi/2, \pm\pi/2)$ points—contrasting the case in \LNOt. Based on our results, we argue that the incommensurate SDW wave vector primarily connects the $\alpha$ pocket and $\beta$ pocket near X/Y points ($Q_2$ in Fig. \ref{fig:motivation}\textbf{b}), rather than most proposed nesting between the $\alpha$ pocket and $\beta$ pocket near $(\pm\pi/2, \pm\pi/2)$ points ($Q_1$ in Fig. \ref{fig:motivation}\textbf{b}). Our findings thus provide key insights into the SDW states in nickelates, offering critical clues to their interplay with superconductivity.

\section{Results}
\LNOf single crystals were synthesized using a vertical optical-image floating-zone furnace with appropriate oxygen pressure \cite{Zhu:2024} ( see Supplementary Materials A for details), exhibiting a characteristic DW transition near 140\,K. Raman scattering measurements were performed using a commercial Raman spectrometer in confocal geometry (see Supplementary Materials B for details). Four polarization configurations were employed: $\hat{x}\hat{x}$, $\hat{x}\hat{y}$, $\hat{x}'\hat{x}'$, and $\hat{x}'\hat{y}'$, where $\hat{x}$ is oriented along the Ni-O-Ni bond and $\hat{x}'$ along the Ni-Ni direction [Fig.~\ref{fig:motivation}(a)].

\LNOf\ belongs to the $C_{2h}$ point group at ambient pressure, which hosts a number of Raman-active phonon modes according to factor group analysis (details can be found in the Supplemental Materials C and Refs.~\cite{Li:2025SB,Gim:2025arXiv,Suthar:2025arXiv,Deswai:2025APL}). All phonon modes appear below 500\,\wn\ in our Raman spectra. The phonon results are discussed in the Supplemental Materials D, while here we focus on the electronic Raman response.  

Figure~\ref{fig:motivation}(c-f) shows the electronic continuum in all four polarization configurations. The continuum exhibits a clear spectral weight loss below 1000\,\wn\ in the $\hat{x}\hat{x}$, $\hat{x}'\hat{x}'$, and $\hat{x}'\hat{y}'$ channels as the temperature decreases from $\approx$ 140 K. The intensity is reduced by up to 50\% in the $\hat{x}\hat{x}$ channel and by approximately 30\% in the $\hat{x}'\hat{x}'$ and $\hat{x}'\hat{y}'$ channels at 500\,\wn. In contrast, the spectral weight remains nearly unchanged in the $\hat{x}\hat{y}$ channel. These results are consistent with previous studies~\cite{Gim:2025arXiv,Suthar:2025arXiv}, demonstrating reproducibility across different sample sources.

\begin{figure}[ht]
   \includegraphics[width=0.4\textwidth]{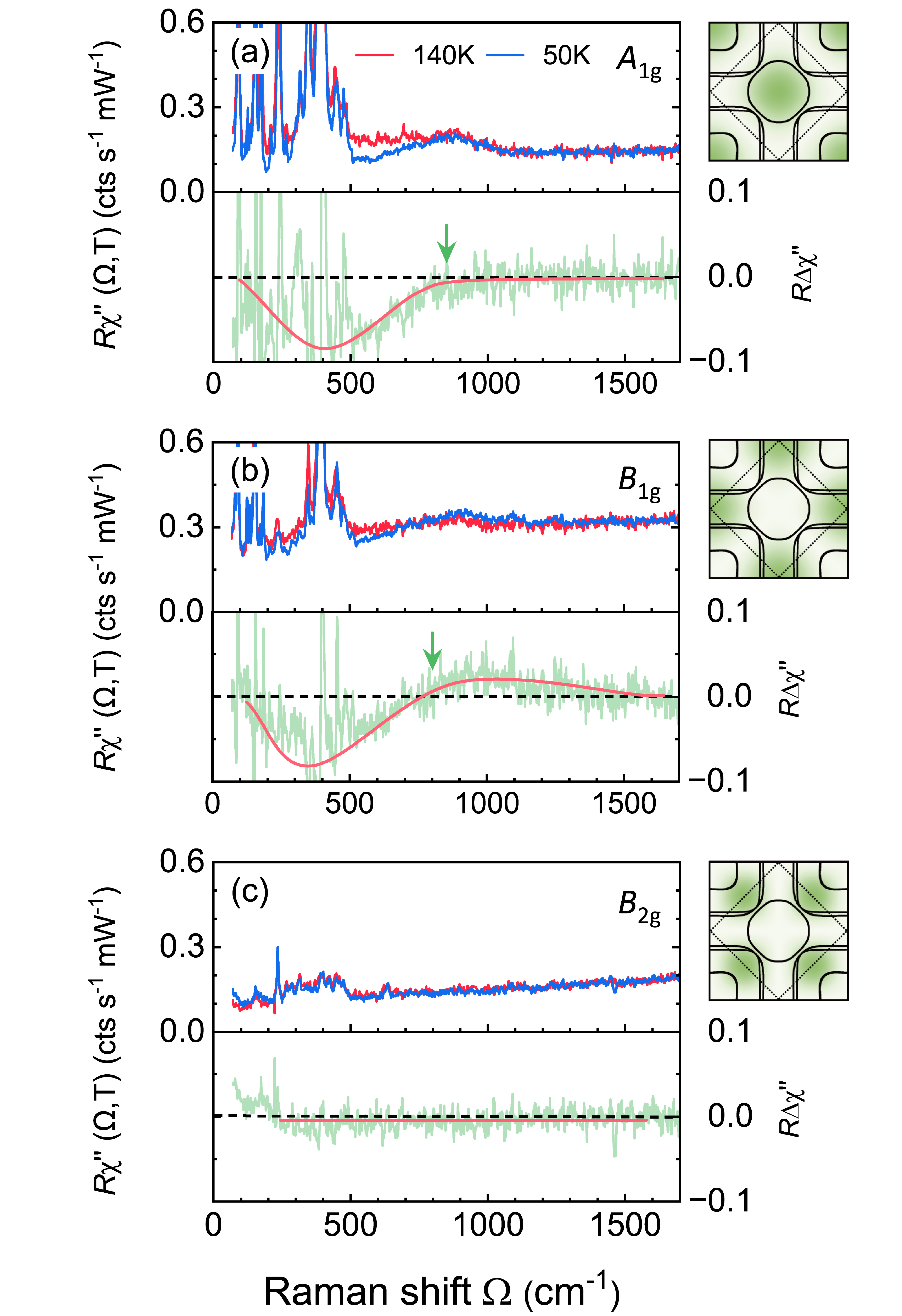}
   \caption{\textbf{Electronic continuum spectra in pure symmetries at 50\,K and 140\,K.} Difference spectra [$\chi''$(50\,K) - $\chi''$(140\,K)] are shown as green curves. Red lines serve as guides to the eye. 
Insets: Crystal harmonics, proportional to the Raman vertices in the first Brillouin zone, superimposed on the Fermi pockets for the \Alg, \Blg, and \BZg\ symmetries.
}

   \label{fig:results}
\end{figure}

Although \LNOf has a monoclinic structure, if we focus solely on the Ni-O plane, as we have done previously in \LNOt~\cite{GeHe:2025} the electronic Raman response can be analyzed using a pseudo-$D_\text{4h}$ symmetry, similar to approaches in cuprates~\cite{Devereaux:2007} and Fe-based superconductors~\cite{Lazarevic:2020}. This framework allows the spectra to be decomposed into three orthogonal symmetries: \Alg, \Blg, and \BZg. The method for isolating spectra in pure symmetries is described in the Supplemental Materials E.  Using this decomposition, we can resolve excitations in different regions of the BZ based on symmetry-imposed selection rules. Following the analogy with tetragonal materials, within the BZ defined for a 1-Ni unit cell, the \Alg channel primarily probes electronic excitations near the BZ center ($\Gamma$ point) and corner (M point). In contrast, the \Blg channel predominantly selects excitations near the BZ boundary around X and Y points, while the \BZg channel is most sensitive to excitations along the diagonal directions of the BZ [see the insets in the top-right corner of Fig.~\ref{fig:results}(a-c) and Supplementary Materials E].

%
%\begin{figure}[ht]
  % \includegraphics[width=0.5\textwidth]{Figure/LNO_Fig3.pdf}
  % \caption{\textbf{Integrat spectral weight in \LNOf.} (a,b) Fits of the electronic continuum at 50 K with phenomenological Drude–Lorentz models for the \Alg\ and \Blg\ spectra, respectively. (c,d) Integrated spectral weight from 0 to 1700\,\wn (red solid circles) as a function of temperature in the \Alg and \Blg channels. The SDW transition temperature is marked by the light green vertical bands.}

  % \label{fig:SW}
%\end{figure}

Figure~\ref{fig:results} presents the Raman spectra in the \Alg, \Blg, and \BZg\ symmetries. In the \Alg channel, a peak near 900\,\wn\ is clearly visible at both 50\,K and 140\,K, with no discernible energy shift. This feature persists even at 300\,K. Suthar \etal have attributed this peak to the opening of an SDW gap~\cite{Suthar:2025arXiv}, although our data suggest a different interpretation (may be a multi-phonon mode or even side effect from the surface). Specifically, we observe instead only a dip feature below 800\,\wn in the difference spectra [$\chi''(50\,\text{K})-\chi''(140\,\text{K})$] of the \Alg channel (see Fig. \ref{fig:results}(a)). In the \Blg channel, the difference spectra exhibit a dip–hump structure with a crossing point near 800\,\wn, consistent with a gap-opening feature, albeit significantly broadened (see Fig. \ref{fig:results}(b)). By contrast, no discernible spectral changes are observed in the \BZg channel across $T_\text{SDW}$ (see Fig. \ref{fig:results}(c)) .

%To illustrate the temperature evolution of the spectral weight (SW), we fitted the continuum using a phenomenological model (see Supplementary Materials F for details). This model is intended only to capture the overall spectral trends and does not carry direct physical significance. From these fits, we extracted the integrated SW to study its temperature dependence. As shown in Fig.~\ref{fig:SW}, the fits (red curves in Fig.~\ref{fig:SW}(a) and (b)) reproduce the continuum in both the \Alg\ and \Blg\ spectra with high fidelity. In the \Alg\ channel, the integrated SW decreases monotonically with cooling, reaching nearly a 20\% reduction at low temperature, with a clear kink near 140\,K [Fig.~\ref{fig:SW}(c)]. In contrast, the \Blg\ channel exhibits a dip in the integrated SW around 130–140\,K [Fig.~\ref{fig:SW}(d)], with a maximum reduction of about 15\%. We associate these anomalies in the temperature-dependent SW with the onset of the SDW transition.

\begin{table}[ht]
\centering
\caption{\textbf{Comparison of SDW features between \LNOf and \LNOt.}}
\label{tab:comparison}

\renewcommand{\arraystretch}{1.5}

\begin{adjustbox}{width=0.45\textwidth}
\begin{tabular}{l c c c l}
\hline\hline
        & \multirow{2}{*}{T$_\text{SDW}$(K)} ~
        & \multicolumn{2}{c}{Q$_\text{SDW}$} 
        & ~\multirow{2}{*}{$\Delta_\text{SDW}$(meV)} \\
\cline{3-4}
        &  & Q vector & BZ &  \\
\hline
\multirow{2}{*}{\LNOf} 
    & \multirow{2}{*}{$\sim$ 140}~ 
    & (0.31,0.31,0)
    & 1-Ni 
    & ~$\Delta_{A_{1g}} \sim 56$ \\
    &  
    & (0,0.61,0) \cite{Zhang2020NC}
    & 2-Ni
    & ~$\Delta_{B_{1g}} \sim 55$ \\
\hline
\multirow{2}{*}{\LNOt} 
    & \multirow{2}{*}{$\sim$ 150} ~
    & (0.25,0.25,0) \cite{RIXS:2024}
    & 1-Ni
    & ~$\Delta_{B_{1g}} \sim 37$ \\
    &  
    & (0,0.5,0) 
    & 2-Ni
    & ~$\Delta_{B_{2g}} \sim 23$ \\
\hline\hline
\end{tabular}
\end{adjustbox}

\end{table}

\section{Discussion}

\begin{figure}[ht]
   \includegraphics[width=0.35\textwidth]{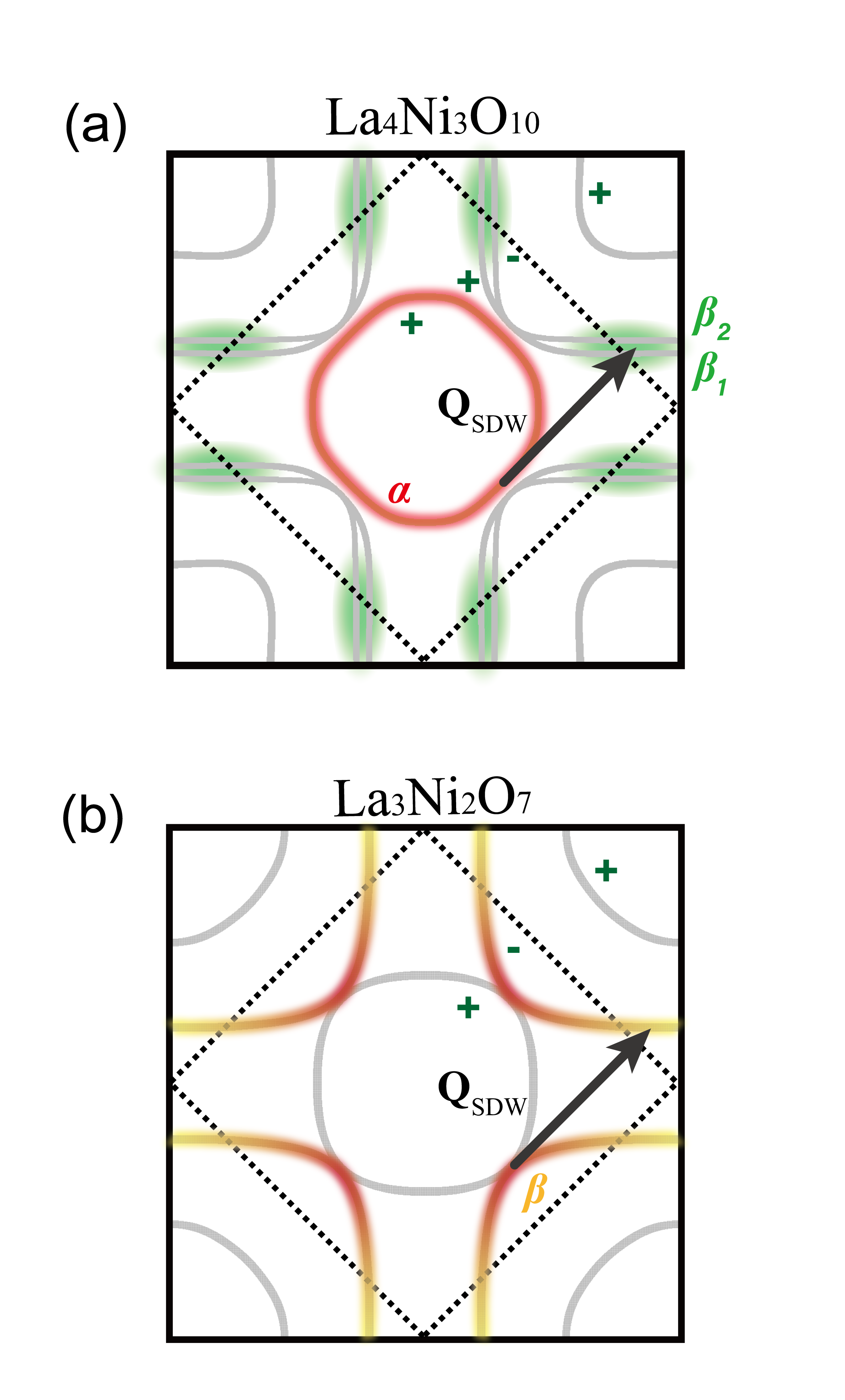}
   \caption{\textbf{Comparison of density-wave gaps and candidate scattering vectors in \LNOf and \LNOt.} Gapped Fermi pockets are highlighted in (a) $\alpha$ pocket and $\beta$ pocket near XY points of \LNOf and (b) $\beta$ pocket of \LNOt. Black arrows indicate possible scattering wave vectors. '+' and '-' signs denote the parity of the bands determined by the layer index (top and bottom Ni-O planes). The colored shading superimposed on the Fermi surface pockets highlights the momentum regions in which energy gaps open.
}
  \label{fig:comparison}
\end{figure}

%Regarding the driving mechanism, our temperature-dependent Raman measurements do not reveal detectable zone-folded phonons across TDW (see Supplementary Materials D), and no sharp collective amplitude mode has been reported so far. Compared with canonical lattice-driven CDW systems, where multiple zone-folded modes and a pronounced amplitude mode are readily resolved in Raman spectra, this lack of clear lattice signatures disfavors a strongly lattice-driven CDW scenario in La4Ni3O10. Our results are instead consistent with an electronically driven density-wave state with a dominant spin component, in line with our La3Ni2O7 Raman study and with STM observations that visualize a CDW modulation likely intertwined with (and possibly secondary to) the underlying SDW order. suggests that lattice distortions across the DW transition are either too subtle to be detected by Raman scattering or, more directly, that a CDW may not be the leading order parameter.

The nature of the DW order in \LNOf has been widely studied~\cite{Lihaoxiang2017NC.8.704,Zhang2020NC,ZhangPhysRevMaterials.4.083402,Li:2025SB,Du2025arXiv.2405.19853v1,WangNanlin:2025,Yuecao:2025arXiv}, with the debate centering on two key issues. The first concerns the driving mechanism of the instability, namely whether it originates from spin, charge, or coupled spin–charge degrees of freedom. The second concerns the detailed nature of the gap opening and the associated ordering wave vector. Regarding the driving mechanism, our temperature-dependent Raman measurements do not reveal detectable zone-folded phonons across $T_{DW}$ (see Supplementary Materials D), and no sharp collective amplitude mode has been reported so far. Compared with canonical lattice-driven CDW systems, such as transition-metal dichalcogenides~\cite{Samnakay:2015,Albertini:2016}, kagome metals~\cite{He:2024}, and rare-earth tritellurides (ReTe$_3$)~\cite{Eiter:2013}, where multiple zone-folded modes and a pronounced amplitude mode are readily resolved in Raman spectra, this lack of clear lattice signatures disfavors a strongly lattice-driven CDW scenario in \LNOf. Instead, our results disfavor a strongly lattice-driven CDW scenario and are consistent with a primarily electronic DW state with a dominant spin component. This conclusion is in line with our previous~\LNOt Raman study~\cite{GeHe:2025} and with STM observations that visualize a CDW modulation likely intertwined with (and possibly secondary to) the underlying SDW order~\cite{Limingzhe:2025}. Following the STM work, a question remains as to whether the SDW ordering wave vector is determined by nesting between the $\alpha$ and $\beta$ pockets or by nesting between parallel segments of the $\beta$ pocket along the $(\pi,\pi)$ direction. In this work, our results provide evidence that favors the $\alpha$–$\beta$ nesting scenario, as discussed in detail below.

%Experimentally, ARPES measurements remain inconclusive regarding whether a gap opens on the $\gamma$ pocket \cite{Lihaoxiang2017NC.8.704, Du2025arXiv.2405.19853v1}, andIndeed, more recent ARPES measurements on \LNOf detected well-defined SDW gap on the $\alpha$ pockets. 

According to the Raman selection rules, the \Alg spectral-weight loss points to a possible gap opening on either the $\alpha$ or $\gamma$ pocket, or both. DFT calculations suggest that a nearly dispersionless flat band on the $\gamma$ pocket dominated by Ni $d_{z^2}$ character is present~\cite{Zhang2024PRB.110.L180501,Yang2024PRB.109.L220506}, thus, making a gap opening on the $\alpha$ pocket a more plausible scenario. The dip–hump feature observed in the \Blg channel suggests a gap opening on the $\beta$ pockets near the X and Y points ($\beta_1$ or $\beta_2$ or both, normally, Raman experiments cannot distinguish them), which has not been reported previously. In contrast, the absence of significant changes in the \BZg channel indicates that the electrons occupying the $\beta$ pockets near the diagonal region of the BZ likely play only a minor role in the SDW formation. Note that Suthar~\textit{et al.} argued that the absence of the gap opening feature in \BZg symmetry is due to the negligible next-nearest-neighbor hopping, resulting in a nearly zero Raman form factor \cite{Suthar:2025arXiv}. In contrast, a pronounced gap feature is clearly observed in the \BZg channel of \LNOt~\cite{GeHe:2025}. Given the fact both of the compounds share quite similar band structure, this comparison suggests that the absence of a gap opening in the \BZg channel in \LNOf is likely due to intrinsic electronic structure properties, where no gap opens in the corresponding region.

These results enable a direct comparison of the SDW properties in \LNOf\ and \LNOt\ (see Tab.~\ref{tab:comparison} and Fig.~\ref{fig:comparison}). In \LNOt, anisotropic gaps open on the $\beta$ pocket with a pronounced coherence peaks \cite{GeHe:2025}. By contrast, in \LNOf, gaps open on the $\alpha$ and $\beta$ pockets near X/Y points ($\beta_1$ or $\beta_2$ or both), but the coherence effect is extremely weak and barely discernible. Specifically, in our previous work, we discussed the possibility that a screening effect in the \Alg channel could have resulted in the failure to observe a gap on the $\alpha$ pocket in \LNOt~\cite{GeHe:2025}. With a direct comparison between \LNOt and \LNOf now available, we have examined the effect of Coulomb screening in both systems using a tight-binding model (see Supplementary Materials G for details). We found that the screening effect reduces the gap excitation intensity by 40\% and 50\% for \LNOf\ and \LNOt, respectively. These comparable reductions indicate that screening alone is insufficient to account for the complete absence of a gap-related excitation in the \Alg symmetry of \LNOt. We therefore infer that $\alpha$ pocket in \LNOt is unlikely to host an SDW gap.
%This calculation is supported by recent optical conductivity experiments where~\LNOf show metallic behavior than ~\LNOt ~\cite{ZheLiu:2024}.

Based on the observed gap-opening characteristics, we propose possible scattering vectors connecting the Fermi pockets. X-ray, neutron scattering, and STM experiments have suggested a characteristic wave vector Q$_1$ (see Fig. \ref{fig:motivation}) that connects the $\alpha$ and $\beta$ pockets \cite{Zhang2020NC,Li2025PRB.112.045132}. Such a scenario would induce a gap on both the $\alpha$ and $\beta$ pockets. In particular, restricted by the mirror symmetry between the top and bottom Ni-O planes in both \LNOt and \LNOf, and considering the interplane antiferromagnetic coupling, the nesting vector would connect bands of opposite parity (marked with '+' and '-' signs in Fig. \ref{fig:comparison}) \cite{Gu:2025, Le：2025}. However, two aspects of this picture conflict with our Raman results. First, there is no gap opening on the $\alpha$ pocket in \LNOt~\cite{GeHe:2025}. Second, we fail to observe a gap opening in the diagonal region of the BZ on the $\beta$ pockets in \LNOf, even though this region satisfies the nesting condition perfectly between the $\alpha$ and $\beta$ pockets. To understand our results, we propose a scattering vector $\mathbf{Q}_{\mathrm{SDW}}$ that connects the $\alpha$ pocket and $\beta$ pocket (near the X/Y points) in \LNOf\ and the $\beta$ pocket itself [between $(\pi/2,-\pi/2)$ and $(\pi,0)$] in \LNOt (Fig.~\ref{fig:comparison}). This configuration naturally accounts for the observed gap openings in both systems. However, it conflicts with the mainstream nesting picture ($Q_1$ in Fig. \ref{fig:motivation}\textbf{b}) discussed above. To reconcile this discrepancy, one would need to consider momentum-dependent scattering and/or strong-coupling mechanism \cite{Liuzhe:2024, ARPESYanglexian：2024, Zhouxingjiang:2024}, which requires further investigation. 

In summary, we have employed polarization-resolved Raman scattering to investigate the density-wave state in \LNOf. By decomposing the spectra into distinct symmetries, we identified spectral-weight anomalies that signal momentum-selective gap openings on the $\alpha$ and $\beta$ pockets. This observation is inconsistent with the commonly proposed scattering vector \textbf{Q}$_1$, and instead supports an alternative vector \textbf{Q}$_2$ connecting the $\alpha$ and $\beta$ pockets near the X/Y points. A comparison with \LNOt further reveals the distinctive momentum selectivity of the SDW order in \LNOt\ and \LNOf, and provides key insights into its connection with superconductivity in layered nickelates.

\noindent\textbf{Note added.} 
During the preparation of our manuscript, we became aware of a recent ARPES study on \LNOf posted on arXiv \cite{Yang:2026}, reporting momentum-selective gap opening. These observations are nearly consistent with our results. 

We thank George Sawatzky, Jiangping Hu, Tao Xiang, Xianin Wu, Ruizhen Huang and Wei Li for fruitful discussions. This work is supported by the National Key Basic Research Program of China (Grants Nos. 2024YFF0727103), the National Natural Science Foundation of China (Grants Nos. 12474473, U23A6015, 12104490, 12375331), D.L.F acknowledges the support by the New Cornerstone Science Foundation (Grant No. NCI202211), and the Innovation Program for Quantum Science and Technology (Grant No. 2021ZD0302803). J.Y. and K.J. acknowledge the support by the CAS Project for Young Scientists in Basic Research (Grant No. 2022YSBR-048). The work at Fudan University was supported by the Key Program of the National Natural Science Foundation of China (12234006), the Innovation Program for Quantum Science and Technology (2024ZD0300100), the National Key R$\&$D Program of China (2022YFA1403202), and the Shanghai Municipal Science and Technology Major Project (2019SHZDZX01).

\bibliography{refs}

@article{Wang:2024,
doi = {10.1088/0256-307X/41/7/077402},
url = {https://dx.doi.org/10.1088/0256-307X/41/7/077402},
year = {2024},
month = {jul},
publisher = {Chin. Phys. Lett.},
volume = {41},
number = {7},
pages = {077402},
author = {Wang, Meng and Wen, Hai-Hu and Wu, Tao and Yao, Dao-Xin and Xiang, Tao},
title = {Normal and Superconducting Properties of $\mathrm{La}_{3}\mathrm{Ni}_{2}\mathrm{O}_{7}$},
journal = {Chin. Phys. Lett.}
}

@article{Sun:2023,
   author = {Sun, Hualei and Huo, Mengwu and Hu, Xunwu and Li, Jingyuan and Liu, Zengjia and Han, Yifeng and Tang, Lingyun and Mao, Zhongquan and Yang, Pengtao and Wang, Bosen and Cheng, Jinguang and Yao, Dao-Xin and Zhang, Guang-Ming and Wang, Meng},
   title = {Signatures of superconductivity near 80 $\mathrm{K}$ in a nickelate under high pressure},
   journal = {Nature},
   volume = {621},
   number = {1},
   pages = {493-498},
   ISSN = {1476-4687},
   DOI = {10.1038/s41586-023-06408-7},
   url = {https://doi.org/10.1038/s41586-023-06408-7},
   year = {2023},
   type = {Journal Article}
}

@article{Li:2025nature,
  author  = {Li, Feiyu and Xing, Zhenfang and Peng, Di and Dou, Jie and Guo, Ning and Ma, Liang and Zhang, Yulin and Wang, Lingzhen and Luo, Jun and Yang, Jie and Zhang, Jian and Chang, Tieyan and Chen, Yu-Sheng and Cai, Weizhao and Cheng, Jinguang and Wang, Yuzhu and Liu, Yuxin and Luo, Tao and Hirao, Naohisa and Matsuoka, Takahiro and Kadobayashi, Hirokazu and Zeng, Zhidan and Zheng, Qiang and Zhou, Rui and Zeng, Qiaoshi and Tao, Xutang and Zhang, Junjie},
  title   = {Bulk superconductivity up to 96 K in pressurized nickelate single crystals},
  journal = {Nature},
  year    = {2026},
  volume  = {649},
  number  = {8098},
  pages   = {871--878},
  doi     = {10.1038/s41586-025-09954-4},
}

@article{Shi2025,
  author   = {Mengzhu Shi and Di Peng and Kaibao Fan and Zhenfang Xing and Shaohua Yang and Yuzhu Wang and Houpu Li and Rongqi Wu and Mei Du and Binghui Ge and Zhidan Zeng and Qiaoshi Zeng and Jianjun Ying and Tao Wu and Xianhui Chen},
  title    = {Pressure induced superconductivity in hybrid Ruddlesden--Popper {La5Ni3O11} single crystals},
  journal  = {Nature Physics},
  year     = {2025},
  volume   = {21},
  number   = {11},
  pages    = {1780--1786},
  issn     = {1745-2481},
  doi      = {10.1038/s41567-025-03023-3},
  url      = {https://doi.org/10.1038/s41567-025-03023-3},
  month    = {11}
}

@article{Zhouxingjiang:2024,
  title = {Orbital-dependent electron correlation in double-layer nickelate $\mathrm{La}_{3}\mathrm{Ni}_{2}\mathrm{O}_{7}$},
  author = {Jiangang Yang and Hualei Sun and Xunwu Hu and Yuyang Xie and Taimin Miao and Hailan Luo and Hao Chen and Bo Liang and Wenpei Zhu and Gexing Qu and Cui-Qun Chen and Mengwu Huo and Yaobo Huang and Shenjin Zhang and Fengfeng Zhang and Feng Yang and Zhimin Wang and Qinjun Peng and Hanqing Mao and Guodong Liu and Zuyan Xu and Tian Qian and Dao-Xin Yao and Meng Wang and Lin Zhao and X. J. Zhou},
  journal = {Nat. Commun.},
  volume = {15},
  issue = {1},
  pages = {4373},
  numpages = {5},
  year = {2024},
  month = {May},
  publisher = {Nature},
  doi = {10.1038/s41467-024-48701-7},
  url = {https://doi.org/10.1038/s41467-024-48701-7}
}

@article{ARPESYanglexian：2024,
doi = {10.1088/0256-307X/41/8/087402},
url = {https://dx.doi.org/10.1088/0256-307X/41/8/087402},
year = {2024},
month = {jul},
publisher = {Chinese Physical Society and IOP Publishing Ltd},
volume = {41},
number = {8},
pages = {087402},
author = {Li, Yidian and Du, Xian and Cao, Yantao and Pei, Cuiying and Zhang, Mingxin and Zhao, Wenxuan and Zhai, Kaiyi and Xu, Runzhe and Liu, Zhongkai and Li, Zhiwei and Zhao, Jinkui and Li, Gang and Qi, Yanpeng and Guo, Hanjie and Chen, Yulin and Yang, Lexian},
title = {Electronic Correlation and Pseudogap-Like Behavior of High-Temperature Superconductor $\mathrm{La}_{3}\mathrm{Ni}_{2}\mathrm{O}_{7}$},
journal = {Chin. Phys. Lett.}
}

@article{ZXSHen:2025,
  title = {Electronic Structure of the Alternating Monolayer-Trilayer Phase of $\mathrm{La}_{3}\mathrm{Ni}_{2}\mathrm{O}_{7}$},
  author = {Abadi, Sebastien and Xu, Ke-Jun and Lomeli, Eder G. and Puphal, Pascal and Isobe, Masahiko and Zhong, Yong and Fedorov, Alexei V. and Mo, Sung-Kwan and Hashimoto, Makoto and Lu, Dong-Hui and Moritz, Brian and Keimer, Bernhard and Devereaux, Thomas P. and Hepting, Matthias and Shen, Zhi-Xun},
  journal = {Phys. Rev. Lett.},
  volume = {134},
  issue = {12},
  pages = {126001},
  numpages = {7},
  year = {2025},
  month = {Mar},
  publisher = {American Physical Society},
  doi = {10.1103/PhysRevLett.134.126001},
  url = {https://link.aps.org/doi/10.1103/PhysRevLett.134.126001}
}

@article{Damascelli:2025,
  title = {Universal electronic structure of layered nickelates via oxygen-centered planar orbitals},
  author = {Christine C. Au-Yeung and X. Chen and S. Smit and M. Bluschke and V. Zimmermann and M. Michiardi and P. Moen and J. Kraan and C. S. B. Pang and C. T. Suen and S. Zhdanovich and M. Zonno and S. Gorovikov and Y. Liu and G. Levy and I. S. Elfimov and M. Berciu and G. A. Sawatzky and J. F. Mitchell and A. Damascelli},
  journal = {arXiv.2502.20450},
  volume = { },
  issue = { },
  pages = { },
  numpages = { },
  year = {2025},
  month = { },
  publisher = { },
  doi = { },
  url = {https://arxiv.org/pdf/2502.20450}
}

@article{Liu2022,
    author = {Liu, Zengjia and Sun, Hualei and Huo, Mengwu and Ma, Xiaoyan and Ji, Yi and Yi, Enkui and Li, Lisi and Liu, Hui and Yu, Jia and Zhang, Ziyou and Chen, Zhiqiang and Liang, Feixiang and Dong, Hongliang and Guo, Hanjie and Zhong, Dingyong and Shen, Bing and Li, Shiliang and Wang, Meng},
   title = {Evidence for charge and spin density waves in single crystals of $\mathrm{La}_{3}\mathrm{Ni}_{2}\mathrm{O}_{7}$ and $\mathrm{La}_{3}\mathrm{Ni}_{2}\mathrm{O}_{6}$},
   journal = {Sci. China Phys. Mech. Astron.},
   volume = {66},
   number = {7363},
   pages = {217411},
   ISSN = {1},
   DOI = {10.1007/s11433-022-1962-4},
   url = {https://doi.org/10.1007/s11433-022-1962-4},
   year = {2022},
   type = {Journal Article}
}

@article{Khasanov:2025,
  author =       {Rustem Khasanov and Thomas J. Hicken and Dariusz J. Gawryluk and Vahid Sazgari and Igor Plokhikh and Loïc Pierre Sorel and Marek Bartkowiak and Steffen Bötzel and Frank Lechermann and Ilya M. Eremin and Hubertus Luetkens and Zurab Guguchia },
  title =        {Pressure-enhanced splitting of density wave transitions in $\mathrm{La}_{3}\mathrm{Ni}_{2}\mathrm{O}_{7-\delta}$},
  publisher =    {Nature},
  journal = {Nat. Phys.},
  volume = {21},
  issue = {10},
  pages = {430-436},
  numpages = {7},
  year = {2025},
  month = {Mar.},

  doi = {10.1038/s41567-024-02754-z},
  url = {https://doi.org/10.1038/s41567-024-02754-z}
}

@article{WutaoNMR:2025,
title = {Pressure-enhanced spin-density-wave transition in double-layer nickelate $\mathrm{La}_{3}\mathrm{Ni}_{2}\mathrm{O}_{7-\delta}$},

author = {Dan Zhao and Yanbing Zhou and Mengwu Huo and Yu Wang and Linpeng Nie and Ye Yang and Jianjun Ying and Meng Wang and Tao Wu and Xianhui Chen},
journal = {Sci. Bull.},
  volume = {70},
  issue = {8},
  pages = {1239-1245},
  numpages = { },
  year = {2025},
  month = { },
  publisher = { },
  doi = {10.1016/j.scib.2025.02.019},
  url = {https://doi.org/10.1016/j.scib.2025.02.019}
}

@article{Yanglexian:2025,
title = {Distinct ultrafast dynamics of bilayer and trilayer nickelate superconductors regarding the density-wave-like transitions},
journal = {Sci. Bull.},
volume = {70},
number = {2},
pages = {180-186},
year = {2025},
issn = {2095-9273},
doi = {https://doi.org/10.1016/j.scib.2024.10.011},
url = {https://www.sciencedirect.com/science/article/pii/S2095927324007503},
author = {Yidian Li and Yantao Cao and Liangyang Liu and Pai Peng and Hao Lin and Cuiying Pei and Mingxin Zhang and Heng Wu and Xian Du and Wenxuan Zhao and Kaiyi Zhai and Xuefeng Zhang and Jinkui Zhao and Miaoling Lin and Pingheng Tan and Yanpeng Qi and Gang Li and Hanjie Guo and Luyi Yang and Lexian Yang}
}

@article{RuiZhou:2025,
	author={Luo, J. and Feng, J. and Wang, G. and Wang, N. N. and Dou, J. and Fang, A. F. and Yang, J. and Cheng, J. G. and Zheng, Guo-qing and Zhou, R.},
	title={Microscopic evidence of charge- and spin-density waves in $\mathrm{La}_3\mathrm{Ni}_2\mathrm{O}_{7-\delta}$ revealed by $\mathrm{^{139}La-NQR}$},
	journal={Chin. Phys. Lett.},
	url={http://iopscience.iop.org/article/10.1088/0256-307X/42/6/067402},
	year={2025},
	
}

@article{Liuzhe:2024,
   author = {Zhe Liu and Mengwu Huo and Jie Li and Qing Li and Yuecong Liu and Yaomin Dai and Xiaoxiang Zhou and Jiahao Hao and Yi Lu and Meng Wang and Hai-Hu Wen},
   title = {Electronic correlations and partial gap in the bilayer nickelate $\mathrm{La}_{3}\mathrm{Ni}_{2}\mathrm{O}_{7}$},
   journal = {Nat. Commun.},
   volume = {15},
   number = {4},
   pages = {7570},
   ISSN = {2041-1723
},
   url = {https://doi.org/10.1038/s41467-024-52001-5},
   year = {2024},
   type = {Journal Article}
}

@article{RIXS:2024,
   author = {Xiaoyang Chen and Jaewon Choi and Zhicheng Jiang and Jiong Mei and Kun Jiang and Jie Li and Stefano Agrestini and Mirian Garcia-Fernandez and Hualei Sun and Xing Huang and Dawei Shen and Meng Wang and Jiangping Hu and Yi Lu and Ke-Jin Zhou and Donglai Feng},
   title = {Electronic and magnetic excitations in $\mathrm{La}_{3}\mathrm{Ni}_{2}\mathrm{O}_{7}$},
   journal = {Nat. Commun.},
   volume = {15},
   number = {4},
   pages = {9597},
   ISSN = {2041-1723},
   url = {https://doi.org/10.1038/s41467-024-53863-5},
   year = {2024},
   type = {Journal Article}
}

@article{Yuxiaohui:2024,
  title = {Density-wave-like gap evolution in $\mathrm{La}_{3}\mathrm{Ni}_{2}\mathrm{O}_{7}$ under high pressure revealed by ultrafast optical spectroscopy},
  author = {Yanghao Meng and Yi Yang and Hualei Sun and Sasa Zhang and Jianlin Luo and Liucheng Chen and Xiaoli Ma and Meng Wang and Fang Hong and Xinbo Wang and Xiaohui Yu},
  journal = {Nat. Commun.},
  volume = {15},
  issue = {2041-1723},
  pages = {10408},
  numpages = {6},
  year = {2024},
  month = {Nov.},
  publisher = {Nature},
  doi = {10.1038/s41467-024-54518-1},
  url = {https://doi.org/10.1038/s41467-024-54518-1}
}

@article{GeHe:2025,
  title = {Anisotropic Electronic Correlations in the Spin Density Wave State of $\mathrm{La}_{3}\mathrm{Ni}_{2}\mathrm{O}_{7}$},
  author = {He, Ge and Shen, Jun and Xie, Shiyu and Zhang, Haotian and Huo, Mengwu and Shu, Jun and Hu, Deyuan and Zhou, Xiaoxiang and Zhang, Yanmin and Qin, Lei and Qiao, Liangxin and Liu, Hengjie and Hu, Chuansheng and Dong, Xijie and Wang, Dengjing and Liu, Jun and Hu, Wei and Yuan, Jie and Yan, Ya-Jun and Qi, Zeming and Jin, Kui and Du, Zengyi and Wang, Meng and Feng, Donglai},
  year = {2025},
  Month = {Aug},
Journal                  = {Research Square},
  doi = {10.21203/rs.3.rs-6952484/v1},
  url = {https://doi.org/10.21203/rs.3.rs-6952484/v1}
}

@article{Devereaux:2007,
   author = {Devereaux, T. P. and Hackl, R.},
   title = {Inelastic light scattering from correlated electrons},
   journal = {Rev. Mod. Phys.},
   volume = {79},
   number = {1},
   pages = {175-233},
   ISSN = {0034-6861},
   DOI = {10.1103/RevModPhys.79.175},
   url = {<Go to ISI>://WOS:000244867600005},
   year = {2007},
   type = {Journal Article}
}

@Article{Deswai:2025APL,
 author = {Deswal, Sonia and Kumar, Deepu and Rout, Dibyata and Singh, Surjeet and Kumar, Pradeep},
    title = {Dynamics of electron–electron correlation and electron–phonon coupled phase progression in trilayer nickelate $\mathrm{La}_{4}\mathrm{Ni}_{3}\mathrm{O}_{10}$},
    journal = {Applied Physics Letters},
    volume = {127},
    number = {7},
    pages = {071903},
    year = {2025},
    month = {08},
    issn = {0003-6951},
    doi = {10.1063/5.0288265},
    url = {https://doi.org/10.1063/5.0288265},
}

@article{Du2025arXiv.2405.19853v1,
  title = {Correlated Electronic Structure and Density-Wave Gap in Trilayer Nickelate $\mathrm{La}_{4}\mathrm{Ni}_{3}\mathrm{O}_{10}$},
  author = {X. Du and Y. D. Li and Y. T. Cao and C. Y. Pei and M. X. Zhang and W. X. Zhao and K. Y. Zhai and R. Z. Xu and Z. K. Liu and Z. W. Li and J. K. Zhao and G. Li and Y. L. Chen and Y. P. Qi and H. J. Guo and L. X. Yang},
  journal = {arXiv.2405.19853},
  volume = { },
  issue = { },
  pages = { },
  numpages = { },
  year = {2024},
  month = { },
  publisher = { },
  doi = { },
  url = {https://arxiv.org/pdf/2405.19853}
}

@article{Gim:2025arXiv,
  title = {Orbital-Selective Quasiparticle Depletion across the Density Wave Transition in Trilayer Nickelate $\mathrm{La}_{4}\mathrm{Ni}_{3}\mathrm{O}_{10}$},
 author = {Gim, Dong-Hyeon and Park, Chung Ha and Kim, Kee Hoon},
  journal = {Phys. Rev. Lett.},
  volume = {135},
  issue = {13},
  pages = {136505},
  numpages = {7},
  year = {2025},
  month = {Sep},
  publisher = {American Physical Society},
  doi = {10.1103/1tp1-y73g},
  url = {https://link.aps.org/doi/10.1103/1tp1-y73g}
}

@Article{Kakoi2024JPSJ.93.053702,
  Title  = {Multiband Metallic Ground State in Multilayered Nickelates $\mathrm{La}_{3}\mathrm{Ni}_{2}\mathrm{O}_{7}$ and $\mathrm{La}_{4}\mathrm{Ni}_{3}\mathrm{O}_{10}$ Probed by 139$\mathrm{La}$-$\mathrm{NMR}$ at Ambient Pressure},
  Author = {Kakoi, Masataka and Oi, Takashi and Ohshita, Yujiro and Yashima, Mitsuharu and Kuroki, Kazuhiko and Kato, Takeru and Takahashi, Hidefumi and Ishiwata, Shintaro and Adachi, Yoshinobu and Hatada, Naoyuki and Uda, Tetsuya and Mukuda, Hidekazu},
  Journal  = {J. Phys. Soc. Jpn.},
  Year = {2024},
  Pages = {053702},
  Volume = {93},
  Comment = {doi: 10.7566/JPSJ.93.053702},
  Doi = {10.7566/JPSJ.93.053702},
  Publisher = {The Physical Society of Japan},
  Url = {https://doi.org/10.7566/JPSJ.93.053702}
}

@Article{Lihaoxiang2017NC.8.704,
  Title                    = {Fermiology and electron dynamics of trilayer nickelate $\mathrm{La}_{4}\mathrm{Ni}_{3}\mathrm{O}_{10}$},
  Author                   = {Li, Haoxiang and Zhou, Xiaoqing and Nummy, Thomas and Zhang, Junjie and Pardo, Victor and Pickett, Warren E. and Mitchell, J. F. and Dessau, D. S.},
  Journal                  = {Nat. Commun.},
  Year                     = {2017},
  Number                   = {1},
  Pages                    = {704--},
  Volume                   = {8},
  ISSN                     = {2041-1723},
  Refid                    = {Li2017},
  Url                      = {https://doi.org/10.1038/s41467-017-00777-0}
}

@Article{Li:2025SB,
  Title                    = {Distinct ultrafast dynamics of bilayer and trilayer nickelate superconductors regarding the density-wave-like transitions},
  Author                   = {Li, Yidian and Cao, Yantao and Liu, Liangyang and Peng, Pai and Lin, Hao and Pei, Cuiying and Zhang, Mingxin and Wu, Heng and Du, Xian and Zhao, Wenxuan and Zhai, Kaiyi and Zhang, Xuefeng and Zhao, Jinkui and Lin, Miaoling and Tan, Pingheng and Qi, Yanpeng and Li, Gang and Guo, Hanjie and Yang, Luyi and Yang, Lexian},
  Journal                  = {Sci. Bull. },
  Year                     = {2025},
  Number                   = {2},
  Pages                    = {180--186},
  Volume                   = {70},
  ISSN                     = {2095-9273},
  Url                      = {https://www.sciencedirect.com/science/article/pii/S2095927324007503}
}

@Article{Li2025PRB.112.045132,
  Title                    = {Direct visualization of an incommensurate unidirectional charge density wave in $\mathrm{La}_{4}\mathrm{Ni}_{3}\mathrm{O}_{10}$},
  Author                   = {Li, Mingzhe and Gong, Jiashuo and Zhu, Yinghao and Chen, Ziyuan and Zhang, Jiakang and Zhang, Enkang and Li, Yuanji and Yin, Ruotong and Wang, Shiyuan and Zhao, Jun and Feng, Dong-Lai and Du, Zengyi and Yan, Ya-Jun},
  Journal                  = {Phys. Rev. B},
  Year                     = {2025},

  Month                    = {Jul},
  Pages                    = {045132},
  Volume                   = {112},

  Doi                      = {10.1103/2p56-xl41},
  Issue                    = {4},
  Numpages                 = {7},
  Publisher                = {American Physical Society},
  Url                      = {https://link.aps.org/doi/10.1103/2p56-xl41}
}

@article{Suthar:2025arXiv,
  Title                    = {Multiorbital character of the density wave instability in $\mathrm{La}_{4}\mathrm{Ni}_{3}\mathrm{O}_{10}$},

  Author                   = {A. Suthar and V. Sundaramurthy and M. Bejas and Congcong Le and P. Puphal and P. Sosa-Lizama and A. Schulz and J. Nuss and M. Isobe and P. A. van Aken and Y. E. Suyolcu and M. Minola and A. P. Schnyder and Xianxin Wu and B. Keimer and G. Khaliullin and A. Greco and M. Hepting},
  journal = {arXiv.2508.06440},
  volume = { },
  issue = { },
  pages = { },
  numpages = { },
  year = {2025},
  month = { },
  publisher = { },
  doi = { },
  url = {https://arxiv.org/pdf/2508.06440}
}

@Article{WangNanlin:2025,
  Title                    = {Origin of the density wave instability in trilayer nickelate $\mathrm{La}_{4}\mathrm{Ni}_{3}\mathrm{O}_{10}$ revealed by optical and ultrafast spectroscopy},
  Author                   = {Xu, Shuxiang and Chen, Cui-Qun and Huo, Mengwu and Hu, Deyuan and Wang, Hao and Wu, Qiong and Li, Rongsheng and Wu, Dong and Wang, Meng and Yao, Dao-Xin and Dong, Tao and Wang, Nanlin},
  Journal                  = {Phys. Rev. B},
  Year                     = {2025},

  Month                    = {Feb},
  Pages                    = {075140},
  Volume                   = {111},

  Doi                      = {10.1103/PhysRevB.111.075140},
  Issue                    = {7},
  Numpages                 = {12},
  Publisher                = {American Physical Society},
  Url                      = {https://link.aps.org/doi/10.1103/PhysRevB.111.075140}
}

@article{Limingzhe:2025,
  title = {Direct visualization of an incommensurate unidirectional charge density wave in $\mathrm{La}_{4}\mathrm{Ni}_{3}\mathrm{O}_{10}$},
  author = {Li, Mingzhe and Gong, Jiashuo and Zhu, Yinghao and Chen, Ziyuan and Zhang, Jiakang and Zhang, Enkang and Li, Yuanji and Yin, Ruotong and Wang, Shiyuan and Zhao, Jun and Feng, Dong-Lai and Du, Zengyi and Yan, Ya-Jun},
  journal = {Phys. Rev. B},
  volume = {112},
  issue = {4},
  pages = {045132},
  numpages = {7},
  year = {2025},
  month = {Jul},
  publisher = {American Physical Society},
  doi = {10.1103/2p56-xl41},
  url = {https://link.aps.org/doi/10.1103/2p56-xl41}
}

@Article{Yang2024PRB.109.L220506,
  Title                    = {Effective model and ${s}_{\ifmmode\pm\else\textpm\fi{}}$-wave superconductivity in trilayer nickelate $\mathrm{La}_{4}\mathrm{Ni}_{3}\mathrm{O}_{10}$},
  Author                   = {Yang, Qing-Geng and Jiang, Kai-Yue and Wang, Da and Lu, Hong-Yan and Wang, Qiang-Hua},
  Journal                  = {Phys. Rev. B},
  Year                     = {2024},
  Month                    = {Jun},
  Pages                    = {L220506},
  Volume                   = {109},
  Doi                      = {10.1103/PhysRevB.109.L220506},
  Issue                    = {22},
  Numpages                 = {7},
  Publisher                = {American Physical Society},
  Url                      = {https://link.aps.org/doi/10.1103/PhysRevB.109.L220506}
}

@Article{Zhang2020NC,
  Title                    = {Intertwined density waves in a metallic nickelate},
  Author                   = {Zhang, Junjie and Phelan, D. and Botana, A. S. and Chen, Yu-Sheng and Zheng, Hong and Krogstad, M. and Wang, Suyin Grass and Qiu, Yiming and Rodriguez-Rivera, J. A. and Osborn, R. and Rosenkranz, S. and Norman, M. R. and Mitchell, J. F.},
  Journal                  = {Nat. Commun.},
  Year                     = {2020},
  Number                   = {1},
  Pages                    = {6003--},
  Volume                   = {11},
  ISSN                     = {2041-1723},
  Refid                    = {Zhang2020},
  Url                      = {https://doi.org/10.1038/s41467-020-19836-0}
}

@Article{Zhang2024PRB.110.L180501,
  Title                    = {${s}^{\ifmmode\pm\else\textpm\fi{}}$-wave superconductivity in pressurized $\mathrm{La}_{4}\mathrm{Ni}_{3}\mathrm{O}_{10}$},
  Author                   = {Zhang, Ming and Sun, Hongyi and Liu, Yu-Bo and Liu, Qihang and Chen, Wei-Qiang and Yang, Fan},
  Journal                  = {Phys. Rev. B},
  Year                     = {2024},

  Month                    = {Nov},
  Pages                    = {L180501},
  Volume                   = {110},

  Doi                      = {10.1103/PhysRevB.110.L180501},
  Issue                    = {18},
  Numpages                 = {9},
  Publisher                = {American Physical Society},
  Url                      = {https://link.aps.org/doi/10.1103/PhysRevB.110.L180501}
}

@Article{Zhang2025PRX.15.021008,
  Title                    = {Bulk Superconductivity in Pressurized Trilayer Nickelate $\mathrm{Pr}_{4}\mathrm{Ni}_{3}\mathrm{O}_{10}$ Single Crystals},
  Author                   = {Zhang, Enkang and Peng, Di and Zhu, Yinghao and Chen, Lixing and Cui, Bingkun and Wang, Xingya and Wang, Wenbin and Zeng, Qiaoshi and Zhao, Jun},
  Journal                  = {Phys. Rev. X},
  Year                     = {2025},

  Month                    = {Apr},
  Pages                    = {021008},
  Volume                   = {15},

  Doi                      = {10.1103/PhysRevX.15.021008},
  Issue                    = {2},
  Numpages                 = {9},
  Publisher                = {American Physical Society},
  Url                      = {https://link.aps.org/doi/10.1103/PhysRevX.15.021008}
}

@Article{Zhang2025PRX.15.021005,
  Title                    = {Superconductivity in Trilayer Nickelate $\mathrm{La}_{4}\mathrm{Ni}_{3}\mathrm{O}_{10}$ under Pressure},
  Author                   = {Zhang, Mingxin and Pei, Cuiying and Peng, Di and Du, Xian and Hu, Weixiong and Cao, Yantao and Wang, Qi and Wu, Juefei and Li, Yidian and Liu, Huanyu and Wen, Chenhaoping and Song, Jing and Zhao, Yi and Li, Changhua and Cao, Weizheng and Zhu, Shihao and Zhang, Qing and Yu, Na and Cheng, Peihong and Zhang, Lili and Li, Zhiwei and Zhao, Jinkui and Chen, Yulin and Jin, Changqing and Guo, Hanjie and Wu, Congjun and Yang, Fan and Zeng, Qiaoshi and Yan, Shichao and Yang, Lexian and Qi, Yanpeng},
  Journal                  = {Phys. Rev. X},
  Year                     = {2025},

  Month                    = {Apr},
  Pages                    = {021005},
  Volume                   = {15},

  Doi                      = {10.1103/PhysRevX.15.021005},
  Issue                    = {2},
  Numpages                 = {11},
  Publisher                = {American Physical Society},
  Url                      = {https://link.aps.org/doi/10.1103/PhysRevX.15.021005}
}

@Article{Zhang2025PRB.111.144502,
  Title                    = {Spin-density wave and superconductivity in $\mathrm{La}_{4}\mathrm{Ni}_{3}\mathrm{O}_{10}$ under ambient pressure},
  Author                   = {Zhang, Ming and Sun, Hongyi and Liu, Yu-Bo and Liu, Qihang and Chen, Wei-Qiang and Yang, Fan},
  Journal                  = {Phys. Rev. B},
  Year                     = {2025},

  Month                    = {Apr},
  Pages                    = {144502},
  Volume                   = {111},

  Doi                      = {10.1103/PhysRevB.111.144502},
  Issue                    = {14},
  Numpages                 = {11},
  Publisher                = {American Physical Society},
  Url                      = {https://link.aps.org/doi/10.1103/PhysRevB.111.144502}
}

@article{Zhu:2024,
  title = {Superconductivity in pressurized trilayer $\mathrm{La}_{4}\mathrm{Ni}_{3}\mathrm{O}_{10-\delta}$ single crystals},
  author = {Zhu, Yinghao and Peng, Di and Zhang, Enkang and Pan, Bingying and Chen, Xu and Chen, Lixing and Ren, Huifen and Liu, Feiyang and Hao, Yiqing and Li, Nana and Xing, Zhenfang and Lan, Fujun and Han, Jiyuan and Wang, Junjie and Jia, Donghan and Wo, Hongliang and Gu, Yiqing and Gu, Yimeng and Ji, Li and Wang, Wenbin and Gou, Huiyang and Shen, Yao and Ying, Tianping and Chen, Xiaolong and Yang, Wenge and Cao, Huibo and Zheng, Changlin and Zeng, Qiaoshi and Guo, Jian and Zhao, Jun},
  journal = {Nature},
  year = {2024},
  volume = {631},
  number = {8021},
  pages = {531--536},
  doi = {10.1038/s41586},
}

@article{Lazarevic:2020,
doi = {10.1088/1361-648X/ab8849},
url = {https://dx.doi.org/10.1088/1361-648X/ab8849},
year = {2020},
month = {jul},
publisher = {IOP Publishing},
volume = {32},
number = {41},
pages = {413001},
author = {Lazarević, N and Hackl, R},
title = {Fluctuations and pairing in $\mathrm{F}$e-based superconductors: light scattering experiments},
journal = {J. Phys.: Condens. Matter}
}

@article{Samnakay:2015,
   author = {Samnakay, R. and Wickramaratne, D. and Pope, T. R. and Lake, R. K. and Salguero, T. T. and Balandin, A. A.},
   title = {Zone-Folded Phonons and the Commensurate–Incommensurate Charge-Density-Wave Transition in $\mathrm{1\mathit{T}-TaSe_2}$ Thin Films},
   journal = {Nano Lett.},
   volume = {15},
   number = {5},
   pages = {2965-2973},
   ISSN = {1530-6984},
   DOI = {10.1021/nl504811s},
   url = {https://doi.org/10.1021/nl504811s},
   year = {2015},
   type = {Journal Article}
}

@article{Albertini:2016,
  title = {Zone-center phonons of bulk, few-layer, and monolayer $\mathrm{1\mathit{T}-TaS_2}$: Detection of commensurate charge density wave phase through $\mathrm{R}$aman scattering},
  author = {Albertini, Oliver R. and Zhao, Rui and McCann, Rebecca L. and Feng, Simin and Terrones, Mauricio and Freericks, James K. and Robinson, Joshua A. and Liu, Amy Y.},
  journal = {Phys. Rev. B},
  volume = {93},
  issue = {21},
  pages = {214109},
  numpages = {7},
  year = {2016},
  month = {Jun},
  publisher = {American Physical Society},
  doi = {10.1103/PhysRevB.93.214109},
  url = {https://link.aps.org/doi/10.1103/PhysRevB.93.214109}
}

@article{He:2024,
  author    = {Ge He and Leander Peis and Emma Frances Cuddy and Zhen Zhao and Dong Li and Yuhang Zhang and Romona Stumberger and Brian Moritz and Haitao Yang and Hongjun Gao and Thomas Peter Devereaux and Rudi Hackl},
  title     = {Anharmonic strong-coupling effects at the origin of the charge density wave in $\mathrm{CsV}_{3}\mathrm{Sb}_{5}$},
  journal   = {Nat. Commun.},
  year      = {2024},
  volume    = {15},
  number    = {1},
  pages     = {1895},
  doi       = {10.1038/s41467-024-45865-0},
  url       = {https://doi.org/10.1038/s41467-024-45865-0},
  issn      = {2041-1723}
}

@article {Eiter:2013,
	author = {Eiter, Hans-Martin and Lavagnini, Michela and Hackl, Rudi and Nowadnick, Elizabeth A. and Kemper, Alexander F. and Devereaux, Thomas P. and Chu, Jiun-Haw and Analytis, James G. and Fisher, Ian R. and Degiorgi, Leonardo},
	title = {Alternative route to charge density wave formation in multiband systems},
	volume = {110},
	number = {1},
	pages = {64--69},
	year = {2013},
	doi = {10.1073/pnas.1214745110},
	publisher = {National Academy of Sciences},
	issn = {0027-8424},
	URL = {https://www.pnas.org/content/110/1/64},
	journal = {Proc. Natl. Acad. Sci.}
}

@article{Gu:2025,
  title = {Effective model and pairing tendency in the bilayer Ni-based superconductor $\mathrm{La}_{3}\mathrm{Ni}_{2}\mathrm{O}_{7}$},
  author = {Gu, Yuhao and Le, Congcong and Yang, Zhesen and Wu, Xianxin and Hu, Jiangping},
  journal = {Phys. Rev. B},
  volume = {111},
  issue = {17},
  pages = {174506},
  numpages = {7},
  year = {2025},
  month = {May},
  publisher = {American Physical Society},
  doi = {10.1103/PhysRevB.111.174506},
  url = {https://link.aps.org/doi/10.1103/PhysRevB.111.174506}
}

@article{Le：2025,
  title = {Opposite-Mirror-Parity Scattering as the Origin of Superconductivity in Strained Bilayer Nickelates},
  author = {Congcong Le and Jun Zhan and Xianxin Wu and Jiangping Hu},
  journal = {arXiv.2501.14665},
  volume = { },
  issue = { },
  pages = { },
  numpages = { },
  year = {2025},
  month = { },
  publisher = { },
  doi = { },
  url = {https://arxiv.org/abs/2501.14665}
}
\end{document}